\documentclass[a4paper,10pt]{article}

\usepackage[utf8]{inputenc}
\usepackage[T1]{fontenc}

\usepackage{a4wide}

\usepackage{amsmath,amssymb}
\usepackage{amsthm}
\usepackage{enumerate} 

\usepackage{authblk}

\usepackage[dvipsnames,table,xcdraw,svgnames]{xcolor}
\definecolor{highlightNEW}{named}{black}
\definecolor{mylilas}{RGB}{170,55,241}
\definecolor{mygreen}{RGB}{70,180,5}
\definecolor{myred}{RGB}{244,63,43}

\usepackage{hyperref}
\hypersetup{
    colorlinks=true,
    breaklinks=true,
    linkcolor=Navy,
    citecolor=Navy,
    urlcolor=Navy
}

\usepackage{graphicx}
\graphicspath{{fig/}}
\usepackage{epstopdf}
\usepackage{subcaption}
\usepackage[section]{placeins} 
\usepackage{rotating}

\usepackage{booktabs}
\usepackage{multirow}

\usepackage[absolute,overlay,showboxes]{textpos}
\TPMargin{2mm}
\textblockrulecolour{Navy}

\usepackage{natbib}

\usepackage{listings}
\newcommand{\doi}[1]{DOI~\href{\detokenize{http://dx.doi.org/#1}}{\detokenize{#1}}}

\newdimen\CdotAxis
\newcommand*{\CdotAux}[3]{%
  {%
    \settoheight\CdotAxis{$#2\vcenter{}$}%
    \sbox0{%
      \raisebox\CdotAxis{%
        \scalebox{#1}{%
          \raisebox{-\CdotAxis}{%
            $\mathsurround=0pt #2#3$%
          }%
        }%
      }%
    }%
    \dp0=0pt %
    \sbox2{$#2\bullet$}%
    \ifdim\ht2<\ht0 %
      \ht0=\ht2 %
    \fi
    \sbox2{$\mathsurround=0pt #2#3$}%
    \hbox to \wd2{\hss\usebox{0}\hss}%
  }%
}
\makeatletter
\def\mathcolor#1#{\@mathcolor{#1}}
\def\@mathcolor#1#2#3{%
  \protect\leavevmode
  \begingroup
    \color#1{#2}#3%
  \endgroup
}
\makeatother

\let\oldalpha\alpha
\renewcommand{\alpha}{\mathcolor{highlightNEW}{\oldalpha}}
\newcommand{\ccode}[2]{\par
        \vspace*{8pt}
        {{\leftskip18pt\rightskip\leftskip
        \noindent{\it #1}\/: #2\par}}\par}
\newcommand{\keywords}[1]{\ccode{Keywords}{#1}}
\newcommand{\email}[1]{\href{mailto:#1}{#1}}

\def\received#1{Received~#1\par}

\def\foliofont{\fontsize{8}{10}\selectfont}
\usepackage[mathscr]{euscript}
\DeclareSymbolFont{rsfs}{U}{rsfs}{m}{n}
\DeclareSymbolFontAlphabet{\mathscrsfs}{rsfs}

\newcommand{\jpTitle}{Robustness and sensitivity analyses for stochastic volatility models under uncertain data structure}
\newcommand{\jpAuthors}{Jan Posp\'{\i}\v{s}il and Tom\'{a}\v{s} Sobotka and Philipp Ziegler}
\newcommand{\jpKeywords}{robustness analysis; sensitivity analysis; stochastic volatility models; bootstrapping; Monte-Carlo filtering}
\newcommand{\jpMSC}{62F35; 62F40; 91G20; 91G70}
\newcommand{\jpJEL}{C52; C58; C12; G12}

\newcommand{\jpDateReceived}{3 March 2017}

\newcommand{\jpDate}{}

\author[1]{Jan Posp\'{\i}\v{s}il\thanks{Corresponding author, \email{honik@kma.zcu.cz}}} 
\author[1]{Tom\'{a}\v{s} Sobotka}
\author[2]{Philipp Ziegler}
\affil[1]{NTIS - New Technologies for the Information Society, Faculty of Applied Sciences, \authorcr University of West Bohemia, Univerzitn\'{\i} 2732/8, 301 00 Plze\v{n}, Czech Republic,}
\affil[2]{Department of Mathematics, University of Rostock, Ulmenstra\ss e 69, 18057 Rostock, Germany}

\ifpdf
\hypersetup{
  pdftitle={\jpTitle},
  pdfauthor={\jpAuthors},
  pdfkeywords={\jpKeywords},
  pdfinfo={
      MSC={\jpMSC},
      JEL={\jpJEL}
  }
}
\fi

\title{\textcolor{Navy}{\textsc{\jpTitle}}}
\date{\jpDate}

\usepackage[absolute,overlay,showboxes]{textpos}
\TPMargin{2mm}
\textblockrulecolour{Navy}

\begin{document}

\maketitle
\begin{textblock}{6}(1.25,1.15)
{\foliofont\noindent This is a pre-print of an article published by Springer Verlag 
in Empirical Economics~57(6), 1935--1958, 2019, \doi{10.1007/s00181-018-1535-3}. \\
Available online \url{https://link.springer.com/article/10.1007/s00181-018-1535-3}.
}
\end{textblock}

\begin{center}
\received{\jpDateReceived}
\end{center}

\begin{abstract}
In this paper we perform robustness and sensitivity analysis of several continuous-time stochastic volatility (SV) models with respect to the process of market calibration. The analyses should validate the hypothesis on importance of the jump part in the underlying model dynamics. Also an impact of the long memory parameter is measured for the approximative fractional SV model. For the first time, the robustness of calibrated models is measured using bootstrapping methods on market data and Monte-Carlo filtering techniques. In contrast to several other sensitivity analysis approaches for SV models, the newly proposed methodology does not require independence of calibrated parameters - an assumption that is typically not satisfied in practice. Empirical study is performed on data sets of Apple Inc. equity options traded in April and May 2015.

\end{abstract}

\ccode{MSC classification}{\jpMSC}
\ccode{JEL classification}{\jpJEL}

\clearpage

\section{Introduction}\label{sec:introduction}

Stochastic volatility (SV) models are common tools for evaluating option prices at financial markets and are of the interest of both academics and practitioners. 
For the applicability in practice, one obtains model parameters by means of calibration or by using filtering estimation techniques as in \cite{Creel15}. We consider calibration to European option prices, as they are widely traded and therefore a sufficient number of contracts is available. As there exists a lot of different SV models, one has to chose an appropriate model for this process. The main assumption of any option pricing model is the structure of modelled dynamics of the underlying. Several empirical studies of various price processes have been analysed in the literature.

Authors \cite{CarrWu03} found the presence of continuous and jump components of modelled market dynamics for the SPX 500 index data. This was done by analysing out-of-the money and at-the-money options' decays in the price for time to maturity reaching zero. As in our case, the authors did not examine prices of the underlying directly which would require extremely high-frequency data sets that could be effected by market microstucture (for time-series tests see e.g. \cite{BarndorffNielsen06, Hwang14}). In \cite{Campolongo06} the use of stochastic volatility models with jumps was recommended because the uncertainty in the estimated option prices mostly came from jump parameters of the considered model. This statement was derived from a study with fictional data and model parameters. We test  the hypotheses of \cite{CarrWu03,Campolongo06} in the case of real market data and we also show that \cite{Campolongo06} method is not suitable for practice (at least for our data sets). A different approach, where a model robustness to varying data structure plays a crucial role, is proposed and applied to real market data sets including Apple Inc. equity options traded in April and May 2015.

In this paper, three subclasses of SV models are considered alongside their representative models: standard \cite{Heston93}, jump-diffusion \cite{Bates96} and so-called approximative fractional jump diffusion (FSV) model, which outperformed the \cite{Heston93} model in terms of in-sample calibration errors in \citep{PospisilSobotka16amf}. Unlike the case of \cite{Bayer15} and many other very recent manuscripts, the FSV model is consider only in the long-memory regime ($H>0.5$). This is due to restrictions on the pricing solution and also in this case we can use the same unifying pricing approach for all three models \citep{BaustianMrazekPospisilSobotka17asmb}. Some comments on the rough volatility regime are to be found in the conclusion.

The considered approaches are tested under uncertainty in the option price structure and are compared with sensitivity and uncertainty analysis tools. \cite{Saltelli04} defined sensitivity analysis as {\it ``the study of how uncertainty in the output of a model (numerical or otherwise) can be apportioned to different sources of uncertainty in the model input''}. We want to know how sensitive calibration errors are with respect to the changing data structure and also how the calibrated parameters are effected. This is done by performing an uncertainty analysis. According to \cite{Saltelli08}, {\it ``uncertainty and sensitivity analyses should be run in tandem, with uncertainty analysis preceding in current practice''}. The method of Sobol indices is the most common approach for global sensitivity analysis. An application of Sobol indices in option pricing can be found in the paper of \cite{Bianchetti15}, were the impact of uncertainty in prices and greeks is measured. However, for real market data one cannot assume independence of the input parameters for calculating Sobol indices. In this paper we use different global sensitivity analysis methods which are discussed e.g. in \cite{Saltelli08}: on a bootstrapped data structure we visualize a dependence of calibrated parameters by scatterplots and hypotheses of jump-presence and long-memory persistence were tested by Monte-Carlo filtering techniques.

Considered models are calibrated from markets (or bootstrapped data) comprising European call options. A European call is a contract that gives the buyer a right to buy a share of the underlying asset for a fixed (strike) price $K$ at some future time $T$. If the buyer observes a stock price at maturity lower than $K$, he or she doesn't utilize her right to buy the asset for $K$. Vice versa, the buyer is exercising the right as long as $S_T\geq K$. This translates into the following pay-off function,
\begin{equation*}
P(x) = \max(x -K,0), \quad \text{where } x = S_T.
\end{equation*}

To answer the question -- what is the fair price to pay for this contract -- one has to build up a set of assumptions on the market that drives $(S_t)_{0<t\leq T}$. Since the Nobel prize winning \cite{BlackScholes73} model one usually considers the stock market to be a stochastic process and the fair price is then obtained using arbitrage-free arguments\footnote{For more detail on the arbitrage pricing see, for instance, \cite{Shreve04}.}. Main differences between the consider models are in the process that drives evolution of the stock price. All approaches in this paper not only assume that the stock price process is of random nature, but also it is assumed that the variance thereof is a stochastic process itself. Hence, a stochastic volatility model can be viewed as a natural extension to the Black-Scholes approach.

Purpose of this article is to help practitioners in the daily calibration process of option pricing models. 
For quantitative tasks beyond the Black-Scholes model, one might face a decision call of choosing a suitable model for the current situation. Different criteria have to be considered, for example the {\it in-sample / out-of-sample errors}, the ability to model the {\it volatility smile} etc. We compare the {\it robustness} of different models with respect to a given option structure. This is important, because the traded equity options can be very different every day, and so their structure (amount of options, strike prices, expiration times, ask-bid-spread, etc.) is another source of uncertainty for the model choice - models might perform differently with respect to different market structures. We show how to analyse this uncertainty, measure it's impact on the predicted option prices and use this as a criteria for choosing a suitable option pricing model. In doing so, we use bootstrapping of the option data and we also introduce several measures of robustness.

The structure of the paper is as follows. In Section \ref{sec:svmodels} we introduce the studied stochastic volatility models and the process of calibration of these models to real market data. In Section \ref{sec:methods} we describe the methodology, in particular the bootstrapping of option prices, as well as we detail the uncertainty and sensitivity analyses. In Section \ref{sec:results} we present the results, we compare all models in terms of variation in model parameters and in bootstrapped option prices. We also provide the results of the Monte-Carlo filtering trials, showing us the importance of the jump intensity for the Bates model and the importance of the long memory parameter for the approximative fractional model. We conclude all obtained results in Section \ref{sec:conclusion}.

\section{Stochastic volatility models}\label{sec:svmodels}

In this paper we focus on a class of stochastic volatility models that captured attention of both practitioners and academics. Modelling approaches of this type
are not restricted by the constant volatility assumption. Many of these models are tractable for a wide range of applications including market calibration that
will be described at the end of this section.

We consider risk-neutral jump-diffusion dynamics of the equity market, where a stock price $S_t$ evolves in time according to the following stochastic
differential equations

\begin{align}
dS_t &= r S_t dt + \sqrt{v_t} S_t d\widetilde{W}^S_t + S_{t-} dJ_t, \label{M1a} \\
dv_t &= p(v_t) dt + q(v_t) d\widetilde{W}^v_t, \label{M2} \\
d\widetilde{W}^S_t d\widetilde{W}^v_t &= \rho\,dt, \quad S_0,~v_0 \in \mathbb{R}^+,  \label{M3}
\end{align}
where $p,q\in C^{\infty}(0,\infty)$ are general coefficient functions for the volatility process and $\rho$ is the correlation between standard Wiener processes
$\widetilde{W}^S_t$ and $\widetilde{W}^v_t$.
To get market dynamics postulated by \cite{Heston93} we  set $dJ_t \equiv 0$, $p(v_t)=\kappa (\theta - v_t)$ and $q(v_t)=\sigma \sqrt{v_t}$.  The set of model
parameters $\Theta^{H}$ is then defined as $\Theta^{H} := \lbrace v_0,~\kappa,~\theta,~\sigma,~\rho \rbrace$.
For the \cite{Bates96} model, functions $p,q$ remain the same only $dJ_t$ corresponds to the compensated compound Poisson process with log-normal jump sizes --
jumps occur with intensity $\lambda$ and their sizes are log-normal with parameters $\mu_J$ and  $\sigma_J$. The set of parameters, in the Bates model case,
consists of 
$\Theta^{B} := \lbrace v_0,~\kappa,~\theta,~\sigma,~\rho,~\lambda,~\mu_J,~\sigma_J \rbrace$. 
Due to more degrees of freedom, the model should provide a better market fit and as was shown in \cite{Duffie00} adding a second jump process to (\ref{M1a})
might not improve the fit any more. Instead of considering a stochastic volatility model with jumps in both underlying and variance dynamics, we use an
approximative fractional process as in \cite{BaustianMrazekPospisilSobotka17asmb, PospisilSobotka16amf}.
Under the approximative fractional model one assumes the same type of jumps as in the previous case, but $p(v_t)= \left[(H-1/2)\psi_t\sigma\sqrt{v} + \kappa
(\theta - v_t)\right]$ and $q(v_t)=\varepsilon^{H-1/2}\sigma \sqrt{v}$, where $\varepsilon>0$ is an approximating factor and $\psi_t$ is an It\^{o} integral:
\begin{equation*}
\psi_t = \int_0^t (t-s+\varepsilon)^{H-3/2} dW^{\psi}_s.
\end{equation*}
The set of parameters $\Theta^{F}:= \lbrace v_0,~\kappa,~\theta,~\sigma,~\rho,~\lambda,~\mu_J,~\sigma_J,~H \rbrace$ also includes the Hurst exponent $H$.
As was shown by \cite{Lewis00} and \cite{BaustianMrazekPospisilSobotka17asmb} respectively, all three models attain a semi-closed form solution not only for plain European options, but
also for more complex structured derivatives - this is crucial for our experiments, a single trial will involve 200 calibrations of each model to different data sets. We also did not perform analyses of models with time dependent parameters which were studied by \cite{MikhailovNogel03}, \cite{Osajima07}, \cite{Elices08}, \cite{Benhamou10} etc. As mentioned in \cite{Bayer15}, the general overall shape of the volatility surface, with respect to equities, does not change in time significantly.

\begin{figure}
\begin{subfigure}[t]{0.5\textwidth}
\centering
\caption{Structure of the Apple Inc. call options (15/5/2015).}
\includegraphics[width=\textwidth]{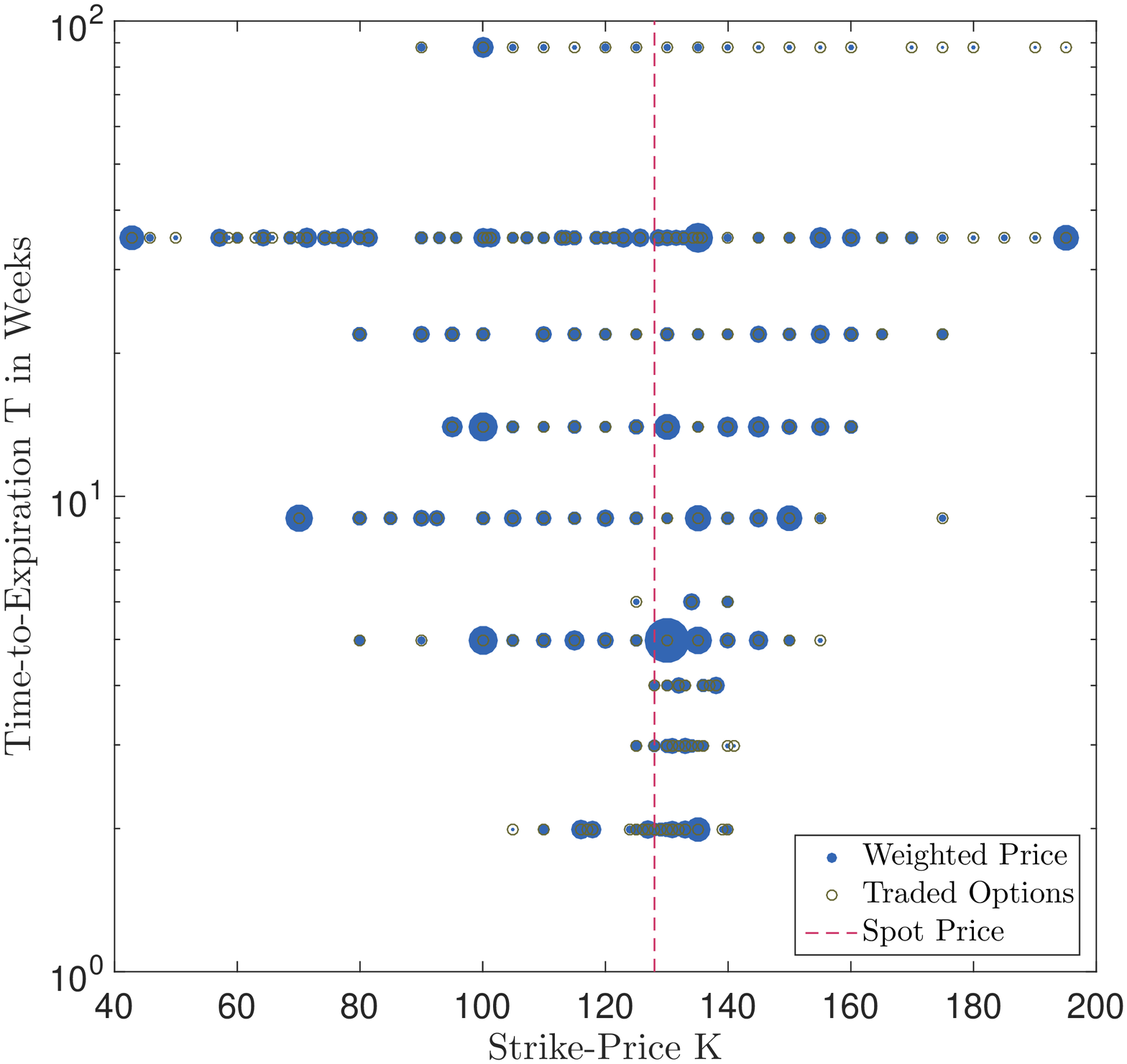}
\end{subfigure}%
\begin{subfigure}[t]{0.5\textwidth}
\centering
{\small
\caption{Parameter bounds for all considered calibration trials.}
\vspace*{0.5cm}
\begin{tabular}{lcc} 
& Lower bound & Upper bound \\
\midrule 
$v_0$ & $0 $ & $1 $ \\
$\kappa$ &$0 $ & $100 $ \\
$\theta$ & $0 $ & $1 $ \\
$\sigma$ & $0 $ & $4 $ \\
$\rho$ & $-1 $ & $1 $ \\
$\lambda$ & $0 $ & $100 $ \\
$\mu_J$ & $-10 $ & $5$ \\
$\sigma_J$ & $0$ & $4 $ \\
$H$ & $0.5 $ & $1 $ \\
\bottomrule
\end{tabular}
\vfill
}
\end{subfigure}
\caption{Data structure and bounds for calibrated parameters. On the left, we depict weighted call prices $w_jC_j^*$ by a ball centred in the $K$ - $T$ plane. The diameter of each filled ball relates to the weighted call price and its center corresponds to the pair $K_j,T_j$.}
\label{Fig:Structure}

\end{figure}

To use the aforementioned models in practice one has to calibrate them to a given market beforehand\footnote{Alternatively one can estimate the parameters from time-series data.}. The calibration process can be viewed as an optimization problem of finding the best fit to the given option price surface.  Let the
surface consist of $N$ options, each with a different strike price ($K$) and time to maturity ($T$) combination. Then typically one uses a utility function defined by the following weighted
least-square criterion:
\[ 
\widehat{\Theta} = \arg \inf\limits_{\Theta} G(\Theta), 
\] 
\begin{equation}\label{Cal1}
G(\Theta)=\sum^N_{j=1}w_j\left( C_j^{\Theta}(T_j,K_j)-C_j^*\right)^2, 
\end{equation}
where $C_j^{\Theta}(T_j,K_j)$ is a model price calculated using the parameter set $\Theta$ and $C^*_j$ represents the $j^{th}$ quoted option price. Weights $
w_j $ are commonly represented as a function of the ask-bid price spread -- based on the results of \cite{MrazekPospisilSobotka16ejor}, we consider the weight function
\begin{equation}
w_j =\frac{1}{\left( C_j^{\text{ask}}-C_j^{\text{bid}}\right)^2}
\end{equation} 
for $j = 1,2, \dots, N$.\\
For the analyses we utilize a data corresponding to European call options on Apple Inc. stock. Four data sets from slightly different time periods
($1/4/2015$, $15/4/2015$, $1/5/2015$ and $15/5/2015$.) and a structure of the newest data set is depicted by Figure \ref{Fig:Structure}. The structure of Apple Inc. stock options should illustrate that we did not restrict our trials to only specific time-to-maturities nor to a specific moneyness range.

\section{Methodology}\label{sec:methods}

In this section we introduce a methodology to analyse a model robustness with respect to uncertain option price structure. This is done by using bootstrapping technique to estimate unobserved samples from the probability density function of the data structure. Also we introduce several measures of robustness that will be later used to compare the models. 

We also define what we mean by a sensitivity analysis. In particular, we would like to analyse whether calibrated values of the jump-intensity parameter $\lambda$ (for the Bates model)  and of the Hurst exponent $H$ (for the FSV model) can significantly effect market fit of the model. As for the measures of robustness, we take advantage of the bootstrapped samples and we use a Monte-Carlo filtering technique to quantify the importance of the mentioned parameters. Both $\lambda$ and $H$ have important consequences for the model choice - by setting $\lambda =0$ and $H=0.5$ we obtain the standard diffusion dynamics postulated by \cite{Heston93}.

\subsection{Bootstrapping Option Prices}\label{subsec:bootstrap} 

The daily option prices are given as a set of ask- and bid-prices with different strike-price $K$ and time to maturity $T$.  This data is new for every day, as the behaviour and the value of the underlying stock, the reference value for the option prices, changes, and it is not statistical in the sense that we only have one dataset for every time instance. Although the option prices usually have some similarities with prices of former days, the focus of traders can change to different options and therefore not all $K \times T$ combinations have to be the same as well as ask- and bid-prices can differ strongly. This can significantly impact calibration results. Therefore we focus on uncertainty in the options structure $K \times T$. Let $X$ be a random variable representing the pair $(K,T)$. Then currently observed strike prices and maturities $(K_j,T_j)$, $j = 1, \dots, N$, as mentioned in (\ref{Cal1}), are samples of $X$ and to each pair we can attach the quoted market option price $C^*_j$. 

To measure the impact of the uncertain option structure on the model calibration, common methods for uncertainty analysis  need statistical data which is not available in practice. In fact, option pricing models are typically recalibrated daily and only to current available and suitable data sets [see e.g. \cite{MikhailovNogel03}, \cite{Yekutieli04}].
In the following we will apply the bootstrapping method to our data set. Since the original paper by \cite{Efron79} and especially his monograph \citep{Efron82}, research activity on the bootstrap method grew dramatically and we refer the reader for example to the book by \cite{Chernick08} and the comprehensive literature review therein. In what follows we apply the standard non-parametric i.i.d. bootstrap method.

We will perform bootstrapping on the set of observed structure $(K_j,T_j)$, $j = 1, \dots, N$, i.e. we obtain a new set ${\bf X}^{\dagger}$ by sampling $N$ times with replacement\footnote{For instance, if $N = 6$ one might obtain ${\bf X}^{\dagger} = (X_2, X_1,X_4,X_4,X_3,X_2)$ where $X_j=(K_j,T_j)$.}. Obviously, for each element of ${\bf X}^{\dagger}$ we can assign a market option price that corresponds to the strike and maturity combination. This provides us with the bootstrap option prices ${\bf C}^{\dagger}=(C_j^{\dagger})_{j=1}^N$. This bootstrap procedure is then repeated $M$ times and hence we get $M$ bootstrap samples ${\bf C}^{\dagger,1},{\bf C}^{\dagger,2},\dots{\bf C}^{\dagger,M}$, each of size $N$. 

Let ${\Theta}^{\dagger,i}$ denote the outcome of the calibration procedure \eqref{Cal1} applied to the bootstrap sample. 
The bootstrap estimate of the mean of the bootstrap replications is 
\begin{equation}
\bar{\Theta} = \frac{1}{M}\sum_{i=1}^M {\Theta}^{\dagger,i}.
\end{equation}

\subsection{Model Comparison}\label{subsec:robustness}

With the bootstrap method, we estimate the calibration parameters $M$ times. As we want to compare different models based
on their robustness, we want to analyze

\begin{itemize}
\item
\dots the variation of the bootstrap replications $\Theta^{\dagger,i}$ and the following
\item
\dots variation of the predicted option prices $C^{\Theta^{\dagger,i}}$, based on the bootstrap replications.
\end{itemize} 

In this paper the comparison of the three SVJD models is widely supported by diagrams for exploring their inner structure and performance. We analyze the bootstrapped calibration parameters with two different diagrams: For the variation in the bootstrapped calibration parameters $\Theta_i$ we use \textit{scatterplot matrices}. For the variation in option prices, we visualize the errors and variations of $C^{\Theta^{\dagger,i}}$ in the $K\times T$-plane.

\subsubsection*{Variation in $\Theta^{\dagger,i}$}\label{subsec:theta_i}

To study the calibration parameter's variation, one can use a variety of methods. First of all, we want to analyze the variation in the bootstrap replications $\Theta^{\dagger,i}$ to derive informations about the model -- e.g. if one really can state that the volatility is fast mean reverting. Additionally, we want to study connections between individual calibration parameters
-- e.g. if the strength of mean reversion varies for different correlations between the Wiener processes. To gather the information, we analyze the variation in $\Theta^{\dagger,i}$ with the help of scatterplot matrices, square matrices with size equal to the number of model parameters. On the diagonal, histograms of the individual calibration parameters are plotted, while the other entries are occupied by scatterplots.

If one further wants to analyze $\Theta^{\dagger,i}$ with statistical methods, normality of the bootstrapped calibration parameters is an important property -- e.g. if we want to calculate confidence intervals for the bootstrap estimate $\bar{\Theta}$ of the calibration parameters. To support such analyses, quantile-quantile plots with respect to a normal distribution are suitable visualization tools.

\subsubsection*{Variation in $C^{\Theta^{\dagger,i}}$}\label{subsec:C^theta_i}

The bootstrapped calibration parameters $\Theta^{\dagger,i}$ contain the uncertainty of the option pricing model with respect to the available options for the calibration. As the bootstrapping was motived by the option pricing structure, vice versa it is helpful to know how this uncertainty affects the model price predictions $C^{\Theta^{\dagger,i}}(K_j,T_j)$ of an individual option $C^*_j$. For this purpose, two measures are introduced:

Firstly, the {\it bootstrap relative error} for the $j$-th option with market price $C^*_j$ is calculated by:
\begin{equation}\label{eq:BRE}
BRE_j =\frac{|\bar{C}_j-C^*_j|}{C^*_j},
\end{equation}
for $j = 1, \dots, N$, where $\bar{C}_j$ is defined as
\begin{equation*}
\bar{C}_j = \frac{1}{M}\sum\limits_{i=1}^{M} C^{\Theta^{\dagger,i}}(K_j,T_j).
\end{equation*}
The measure indicates an individual price prediction error of the bootstrap estimation $\bar{C}$ normalized with the market option price. Using bootstrap relative errors, we should be able to detect systematic prediction errors, which can come from the specific option pricing structure.

We are also interested in the variance of prediction error $|C^{\Theta^{\dagger,i}}(K_j,T_j)-C^*_j|$ for the bootstrapped parameters $\Theta^{\dagger,i}$ with respect to bootstrap trials $i = 1, \dots, M$. To be able to compare variances of predictions for options with different prices $C^*_j$, we use relative errors as before to get the variance error measure $V_j$ for the $j$-th option:
\begin{equation}\label{eq:BRVar}
V_j = \operatorname{Var} \left(\frac{|C^{\Theta^{\dagger,i}}(K_j,T_j)-C^*_j|}{C^*_j}\right),
\end{equation}
This measure is evaluated for all traded options $j=1,\dots,N$.

The error and variance measures are visualized with diagrams in the $K \times T$ plane. Each traded option is marked with a circle which is centred according to $(K, T)$ of the contract. For a clear arrangement, the $T$-axis is in logarithmic scale, because there are many traded options with short but slightly different time to maturity. Current asset price, the reference for the option prices, is plotted as a dashed line. Finally, the average relative error respectively the variance of the
bootstrapped prices are visualized as balls, where the balls area is scaled with the error/variance.

\subsection{Sensitivity analysis} \label{sec:sensitivity_method}

According to \cite{Saltelli08}, the scatterplots can be used as a tool for sensitivity analysis to measure the impact of input parameters on the model output.
Additionally, in this paper we would like to know, if fractionality of stochastic volatility and jumps are important for the robustness of option pricing models. Fractionality is represented by the Hurst parameter $H$, while jumps are represented by the intensity parameter $\lambda$ (which is linked to the parameters $\sigma_J$ and $\mu_J$). Therefore, the importance of jumps and fractionality can be translated into the question, if $H$ and $\lambda$ have an impact on the quality of the calibration result. This question will be addressed by the Monte-Carlo filtering technique, which analyzes if a distribution of values of a chosen parameter effects significantly some specific quality measure.\\
In our context, we have chosen the following Monte-Carlo filtering technique\footnote{For more details on Monte-Carlo filtering approaches see, for instance \cite{Saltelli08}.}: To each set of
calibrated model parameters, obtained from the bootstrapped data, we assign average absolute relative error (AARE) with respect to the whole set of traded
options as a quality measure for the parameter set. This enables us to divide the sets of parameters into a behavioural (well fitting) group and a non-behavioral (poor fitting) group with respect to the AARE measure. As a behavioural set of parameters we consider parameters for which AARE is in
the lower $3/8$ quantile. A non-behavioural set, on the other hand, consists of parameter sets that lead to the worst $37.5 \%$ of the AARE values (upper $3/8$
quantile). The rest ($1/4$ of the results) is consider as a ``grey zone'' and is not taken into account for the comparison\footnote{In this case, we will not be
able to decide whether the parameters lead to a good or bad description of the modelled market}. For the considered parameter set groups we perform a two-sample
Kolmogorov-Smirnov test to verify the null hypothesis whether both sets are sampled from the same distribution function. According to \cite{Saltelli08}, by
rejecting the null hypothesis at a reasonable level of significance\footnote{For all trials we use ``standard'' $\alpha = 5\%$ level of significance. In most of
the trials we could have even lower $\alpha$ and still we would reject the null hypothesis.} we show that the parameters are important with respect to the
calibration procedure. However, if we are not able to reject the hypothesis then we cannot judge the importance of the selected parameter.

For the Bates model we would like to answer whether the jumps are worth implementing to fit the observed market or if one should stay within the Heston model
framework ($\lambda \neq 0$). We also judge the importance of the Hurst parameter in the fractional stochastic volatility case. 

\section{Results}\label{sec:results}

In our trials, the bootstrap calibration was performed $M = 200$ times. In the following text we discuss the results based on the four data sets mentioned above\footnote{All results and data are available in supplementary materials.}. For example the data set from $15^{th}$ May consists of 197 options and the most weight is typically assigned to the at-the-money contracts allocated near spot price in Figure \ref{Fig:Structure}. Apart from that, weights are almost evenly distributed in the $K\times T$ plane.

We start the model comparison by examining the overall calibration errors of all three
models as seen in Table \ref{Tab:Overallerror}. We note that the additional model features of the Bates model (jumps) and the FSV model (jumps and approximative fractional Brownian motion) usually lead to a better market fit. The best average relative errors were obtained on the 15/4/2015 data set (FSV model reached $2.16\%$ error) and the worst market fit in terms of the consider measure was w.r.t. 1/5/2015 data and the Heston model ($6.58\%$). We conclude that using the Heston model we were able to retrieve similar error measures to the Bates and FSV model only for the data set from 15/5/2015.

\begin{table}[hb]
\centering
{\small
\caption{Overall calibration errors of the three models for Apple Inc. stock on all four datasets.}
\label{Tab:Overallerror}
\begin{tabular}{lllll} 
Trading day & 1/4/2015 & 15/4/2015 & 1/5/2015 & 15/5/2015 \\
\midrule 
Heston model & $5.15\%$ & $3.79\%$ & $6.58\%$ & $3.39\%$\\
Bates model & $3.73\%$ & $3.57\%$ & $5.77\%$ & $3.41\%$\\
FSV model & $2.21\%$ & $2.16\%$ & $5.89\%$ & $3.20\%$\\
\bottomrule
\end{tabular}
\vfill
}
\end{table}

\subsection*{Variation in $\Theta^{\dagger,i}$}

\begin{sidewaysfigure}
\centering%
\includegraphics[width=\textwidth]{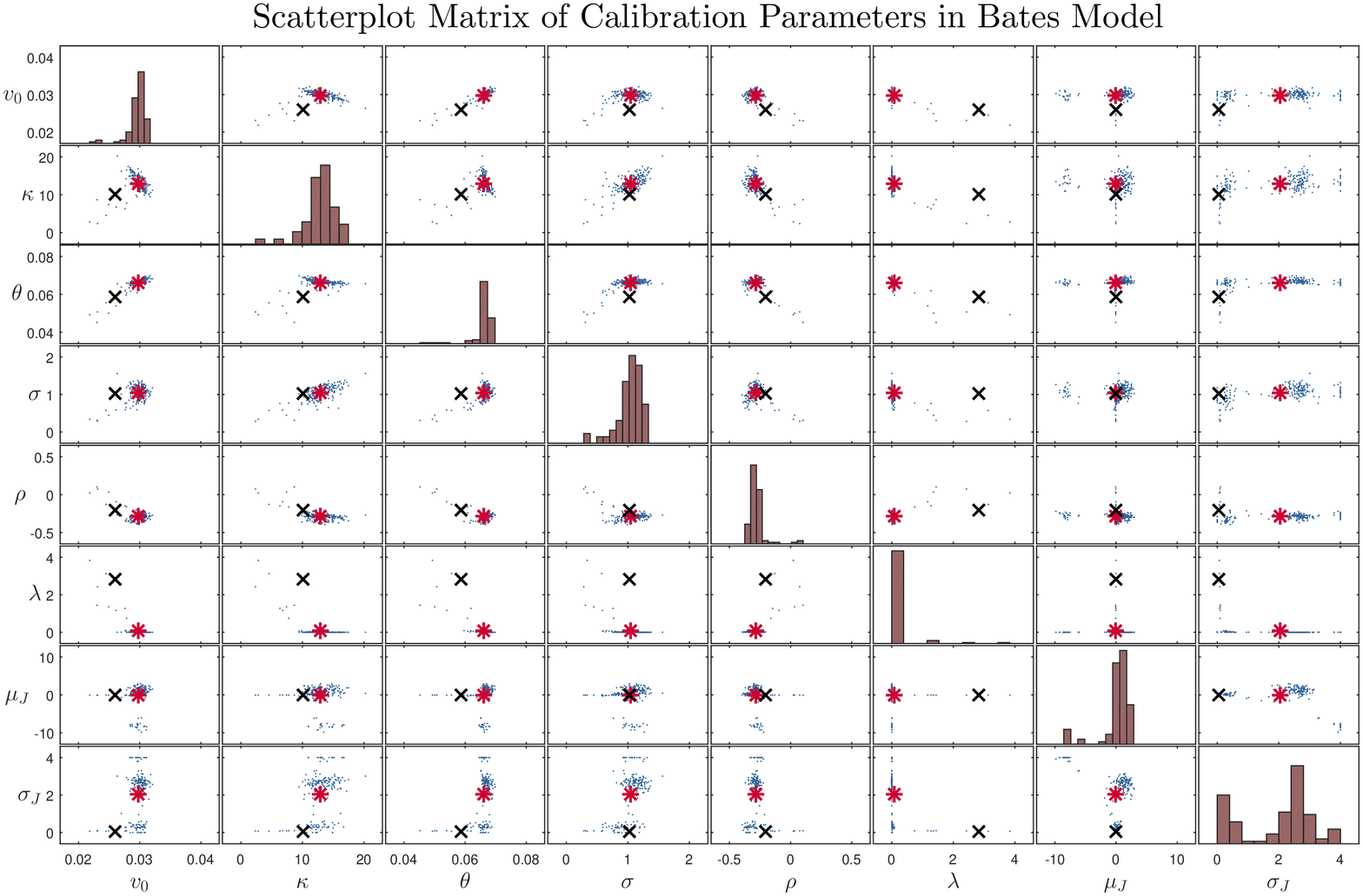}
\centering
\caption{Scatterplot matrix for the Bates model (15/5/2015, for results on other data sets see the supplementary materials). Diagonal elements depict histograms of parameter values obtained by bootstrap calibrations (e.g. the fist histogram corresponds to the values of $v_0$). Off-diagonal elements illustrate a dependence structure for each parameter pair. In those figures, a black cross represents the reference value of the specific parameter (obtained from calibration to the whole data set) and by a red star we depict the bootstrap estimate of the value - the closer the two are, the better.}
\label{Fig:ScatterBates}
\end{sidewaysfigure}

\begin{sidewaysfigure}
\includegraphics[width=\textwidth]{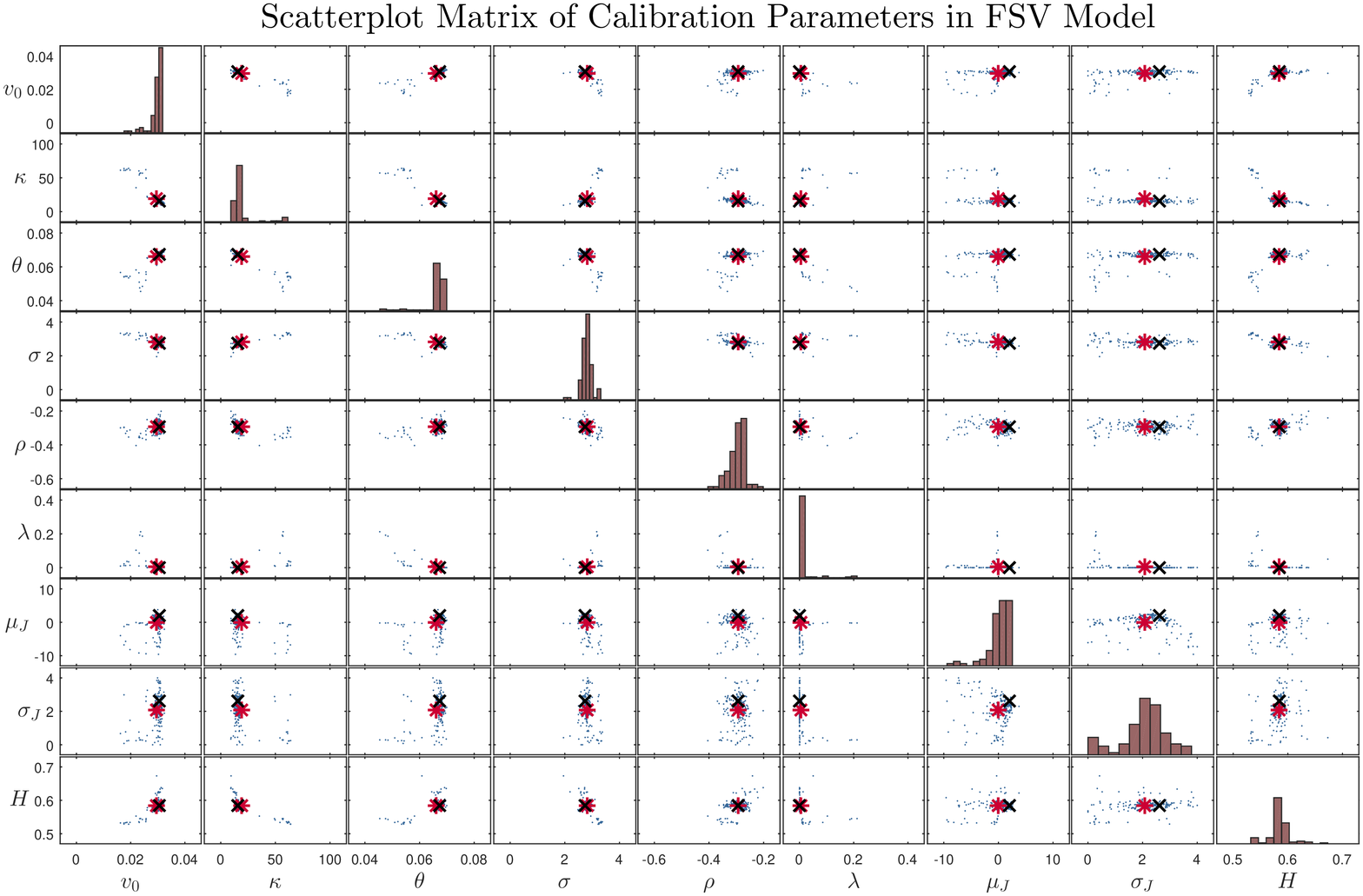}
\centering
\caption{Scatterplot matrix for the FSV model (15/5/2015, for results on other data sets see the supplementary materials). Diagonal elements depict histograms of parameter values obtained by bootstrap calibrations (e.g. the fist histogram corresponds to the values of $v_0$). Off-diagonal elements illustrate a dependence structure for each parameter pair. In those figures, a black cross represents the reference value of the specific parameter (obtained from calibration to the whole data set) and by a red star we depict the bootstrap estimate of the value - the closer the two are, the better.}
\label{Fig:ScatterFSV}
\end{sidewaysfigure}

In Figures \ref{Fig:ScatterBates} and \ref{Fig:ScatterFSV}, the scatterplot matrices of the parameters $\Theta^{\dagger,i}_{\textnormal{Bates}}$ and $\Theta^{\dagger,i}_{\textnormal{FSV}}$ are depicted (the results
for $\Theta^{\dagger,i}_{\textnormal{Heston}}$ are similar to the ones for the Bates model, we discuss them shortly at the end of the section). First and foremost, we inspect if we  reached the lower and upper bounds for the calibration parameters (the bounds are listed in Figure \ref{Fig:Structure}b). This can indicate

\begin{itemize}
\item \dots for a zero bound (e.g. $\kappa \geq 0$), that a model parameter (e.g. mean reversion $\kappa$) could be dropped,
\item \dots for non-zero bounds (e.g. $\kappa \leq 100$), that they should be reselected if it is not in contradiction with the parameter interpretation and  if it does not breach model restrictions.
\end{itemize}

For correlation $\rho$, the natural limits at $-1$ and $1$ indicate that only one Brownian motion can model both, the random movement of the asset
price and its volatility. 
Additionally, $\rho$ and $\mu_J$ include zero in the
interior of their calibration range, which should be considered during the exploration of scatterplots. E.g. for the uncorrelated Heston model \cite{MarcoMartini10} showed an explicit formula which not even needs numerical integration and, possibly, even the other models might be simplified. On the contrary, the value $\mu_J=0$ has no model reducing consequences. Last but not least, a dependence structure between the calibration parameters can be obtained from a single scatterplot and we are able to compare the bootstrap mean $\bar{\Theta}$ (red star) and the parameters $\Theta$ from the overall calibration (black cross).

Starting with the Bates model and the last criteria, one cannot observe a noticeable accumulation of $\rho$ and $\mu_J$ at zero in the histograms at the diagonal of Figure \ref{Fig:ScatterBates}. Further on, the histograms show that the parameters $v_0,\kappa,\theta,\sigma,\rho$ and $\mu_J$ have no
concentration at their limits. However, the small values of $\lambda$ (mostly between $10^{-3}$ and $10^{-4}$) and the accumulation of $\sigma_J$ at zero are noticeable. Moreover, if one looks at the scatterplot between $\lambda$ and $\mu_J$, $\lambda$ and $\sigma_J$, one observes that {\it either} $\lambda$  is nearly zero {\it or} $\sigma_J$ and $\mu_J$ are close to zero. For the model this means, we have two possible cases: either the Bates model has very rare jump, or it has many very small jumps of the same intensity. If the jump-frequency $\lambda$ approaches zero, then the average jumps sizes $\mu_J$ are almost for all calibrations negative. This statistical connection between $\lambda$, $\mu_J$ and $\sigma_J$ should be considered at the calibration by a general modelling decision. One option would be to fix the jump intensity parameter $\lambda$ beforehand.

Furthermore, the scatterplots depict that from all parameters $\kappa$ is the one with the most correlations. One can see from the scatterplots that the
stronger the mean reversion is,

\begin{itemize}
\item \dots the higher is the volatility of volatility $\sigma$,
\item \dots the lower is the initial volatility $v_0$ and the long-run volatility $\theta$,
\item \dots the more negative the correlation $\rho$ between the two Brownian motions is.
\end{itemize}

The difference between $\bar{\Theta}_{\textnormal{Bates}}$ and $\Theta_{\textnormal{Bates}}$ for most parameters is not very large, apart from $\lambda$ and $\sigma_J$: the overall calibration
resulted in a model with many small jumps of the same size, while the bootstrap resulted in the mean in a model with rare jumps of different size.

In Figures \ref{Fig:ScatterBates} and \ref{Fig:ScatterFSV} we can see that the histograms of FSV and Bates model are quite similar for $v_0$, $\theta$, $\rho$,
$\lambda$, $\mu_J$ and $\sigma_J$. The rate of mean-reversion $\kappa$ in FSV model is slightly different to Bates model, as in some of the trials we can get
a fast mean-reversion rate, $\kappa \gtrapprox 50$. Furthermore, $\sigma$ in FSV model is significantly higher than in Bates model. This can be explained by the
scaling with $\varepsilon^{H-1/2}$ (see equation \eqref{M2} and the definition of $q(v_t)$). The Hurst parameter $H$ is positively correlated with $v_0$, $\theta$ and
$\rho$ and negatively with $\kappa$ and $\sigma$. In Tables \ref{Tab:CorrBates} and \ref{Tab:CorrFSV} we provide pairwise correlation coefficients for both models.
Note that these are all stochastic volatility parameters, a connection of the Hurst parameter with the jump
parameters is not obvious in the scatterplots, see also Table \ref{Tab:CorrFSV}. Finally, in FSV model $\bar{\Theta}_{\textnormal{FSV}}$ and $\Theta_{\textnormal{FSV}}$ are very close together which is a desirable result.

\begin{table}[h]
\centering
{\small
\caption{Pairwise correlation coefficients of calibrated parameters for the Bates model and the data from 15/5/2015.}
\label{Tab:CorrBates}
\begin{tabular}{lcccccccc} 
           & $v_0$ & $\kappa$ & $\theta$ & $\sigma$ & $\rho$ & $\lambda$ & $\mu_J$& $\sigma_J$ \\
 \midrule 
$\lambda$  & $-0.7456$ & $-0.0356$ & $-0.7940$ & $-0.5483$ & $~0.7235$ & $--$ & $~0.0012$& $-0.2984$ \\
$\mu_J$    & $~0.0866$ & $-0.0356$ & $~0.0938$ & $~0.0526$ & $-0.1065$ & $~0.0012$ & $--$& $-0.3553$ \\
$\sigma_j$ & $~0.2647$ & $~0.3807$ & $~0.3726$ & $~0.5504$ & $-0.0207$ & $-0.2984$ & $-0.3553$& $--$ \\
 \bottomrule
\end{tabular}
\vfill
}
\end{table}

For the Heston model, the scatterplot matrix showed similar results as the upper left $5\times5$ submatrix of Figure~\ref{Fig:ScatterBates}. 
There were no accumulations of $\Theta^{\dagger,i}_{\textnormal{Heston}}$ at the bounds and the correlation seemed nearly linear. Independence of the calibration
parameters, necessary for the sensitivity analysis method proposed in \cite{Campolongo06}, cannot be assumed for any of the models. Thus, this method is
not suitable in our context.

\begin{table}[h]
\centering
{\small
\caption{Pairwise correlation coefficients of calibrated parameters for the FSV model and the data from 15/5/2015.}
\label{Tab:CorrFSV}
\begin{tabular}{lccccccccc} 
           & $v_0$ & $\kappa$ & $\theta$ & $\sigma$ & $\rho$ & $\lambda$ & $\mu_J$& $\sigma_J$ &$H$ \\
 \midrule 
$\lambda$  & $-0.4622$ & $~0.5778$ & $-0.7648$ & $~0.3080$ & $-0.3574$ & $--$ & $-0.0637$& $-0.3983$ & $-0.3600$ \\
$\mu_J$    & $~0.3347$ & $-0.3603$ & $~0.3055$ & $-0.3131$ & $-0.0725$ & $-0.0637$ & $--$& $-0.1080$ &$~0.2063$ \\
$\sigma_j$ & $~0.4027$ & $-0.3974$ & $~0.4286$ & $-0.1420$ & $~0.0986$ & $-0.3983$ & $-0.1080$& $--$ &$~0.2894$ \\
$H$        & $0.7134$  & $-0.7859$ & $~0.6600$ & $-0.6059$ & $~0.4673$ & $-0.3600$ & $~0.2063$& $0.2894$ &    $--$\\
 \bottomrule
\end{tabular}
\vfill
}
\end{table}

\begin{figure*}
\centering%
\includegraphics[width=\textwidth]{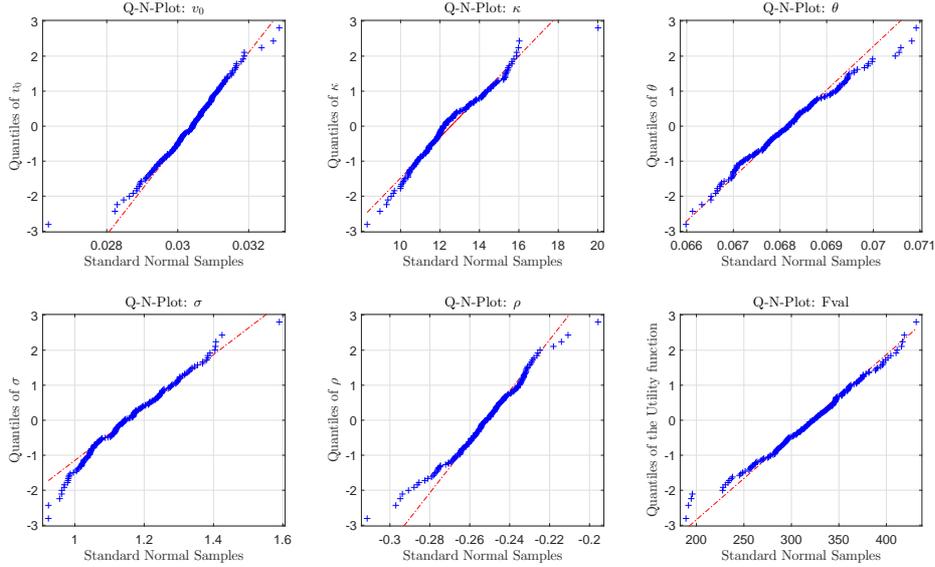}
\caption{Q-N plots for Heston model parameters $(\nu_0,\kappa,\theta,\sigma,\rho)$ and corresponding values \textit{Fval} of the calibration utility function (15/5/2015).}
\label{Fig:QQHeston}
\end{figure*}

One can notice from the Q-N plots in Figure \ref{Fig:QQHeston}, that normality of the bootstrapped parameters and of the resulting calibration error can be
assumed for the Heston model -- for Bates and FSV this was not the case. Therefore, statistical techniques which assume normality of the data could be used to
further analyze the Heston model, but not for comparison of all three models.

\subsection*{Variation in $C^{\Theta^{\dagger,i}}$}

\begin{figure*}
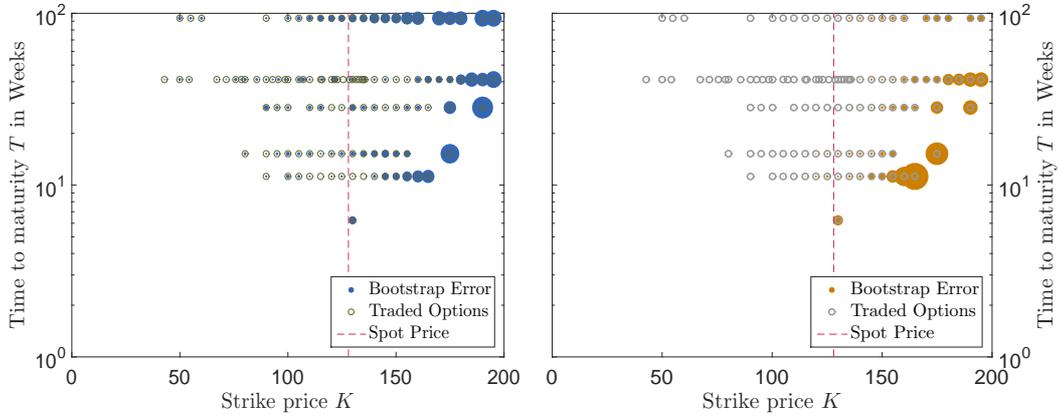
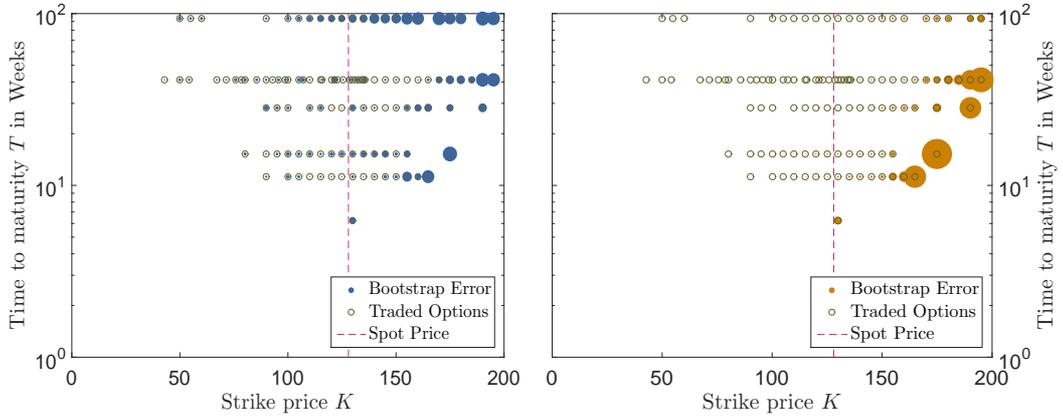
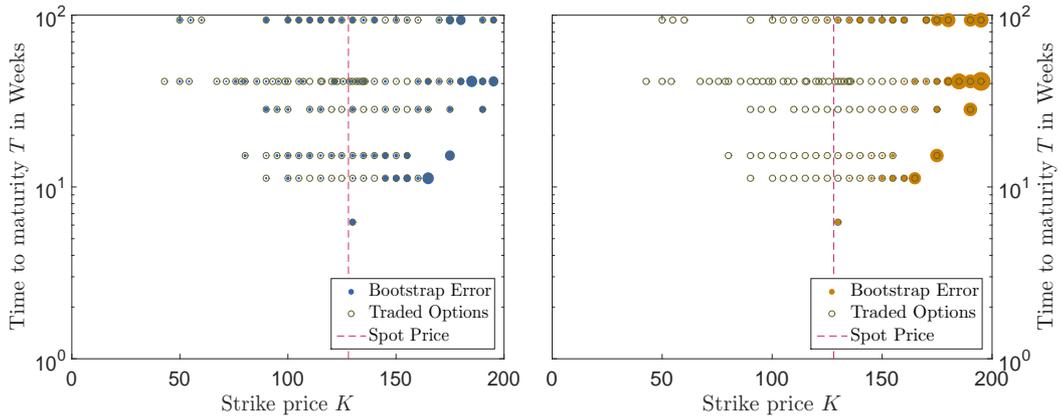

\begin{subfigure}[b]{\textwidth}
\centering
\includegraphics[width=0.45\textwidth,height=55mm]{fig/ScatExpHeston-4-01.eps}\quad
\includegraphics[width=0.45\textwidth,height=55mm]{fig/ScatVarianceHeston-4-01.eps}
\caption{Heston model}
\end{subfigure}

\begin{subfigure}[b]{\textwidth}
\centering
\includegraphics[width=0.45\textwidth,height=55mm]{fig/ScatExpBates-4-01.eps}\quad
\includegraphics[width=0.45\textwidth,height=55mm]{fig/ScatVarianceBates-4-01.eps}
\caption{Bates model}
\end{subfigure}

\begin{subfigure}[b]{\textwidth}
\centering
\includegraphics[width=0.45\textwidth,height=55mm]{fig/ScatExpFSV-4-01.eps}\quad
\includegraphics[width=0.45\textwidth,height=55mm]{fig/ScatVarianceFSV-4-01.eps}
\caption{FSV model}
\end{subfigure}
\caption{Calibration errors and variance of the obtained option price surface in $K \times T$ plane. On the left, we depict bootstrap relative errors $BRE_j$ for all bootstrap calibrations w.r.t. 
1/4/2015 data set by filled circles with diameter proportionate to the error. On the right, the variance $V_j$ of each option price is illustrated - as before, a diameter of a specific filled circle is proportionate to the option price variance. For results on other data sets see the supplementary materials.}
\label{Fig:C_Theta_pic}
\end{figure*}

In Figure \ref{Fig:C_Theta_pic}, the bootstrap relative errors \eqref{eq:BRE} and the bootstrap variances \eqref{eq:BRVar} for every call option are shown. The structure of
the errors appears similar for all three models -- a result that fits very good to the overall calibration errors for the considered data set. The highest
errors appear for all three models for the OTM options, especially if the strike price is grater than 150 USD. For all data sets the lowest values of the bootstrap relative  error were obtained by the FSV model. 

The variance error measure $V_j$ shows for all three models the same structure in the $K\times T$ plane, but values differ strongly. 
In the data set from 1/4/2015, the measured variances were the lowest for the FSV model again, whereas the Bates model provided us with the worst values. 
Surprisingly, Bates model was clearly outperformed for all considered data sets. On the other hand results for Heston and FSV slightly differed for the other data sets. 
We refer a reader to the corresponding figures in supplementary materials.

\subsection{Sensitivity analysis}

In this section we would like to inspect parameter reducing possibilities for the Bates and the FSV model. We check whether the jump-intensity $\lambda$ plays a crucial role in obtaining good error measures for the Bates model calibration. If we fix $\lambda =0$ we would obtain the standard Heston model. Similarly we proceed with the FSV model, where we inspect if we can profit from setting $H>0.5$, unlike formally fixing $H=0.5$ to obtain the Bates model.

\subsection*{Importance of jumps}

For all available datasets we managed to reject the null hypothesis that both the behavioural and non-behavioural sets of $\lambda$ are from the same
distribution with $5 \%$ level of significance. In Table \ref{Tab:SenLambda} we also display the p-values obtained from the 2-sample Kolmogorov-Smirnov test. These
are the maximal levels of significance that would lead to not rejecting the null hypothesis. Hence, we are
able to conclude the similar result as in \cite{Campolongo06} - the jump term is of significant help for calibration trials. In our case the conclusion is drawn from the real market data and using the Monte-Carlo filtering technique introduced in Section \ref{sec:methods}. However, it is worth mentioning that this technique identifies input parameters which influence extremes in the output (quality of calibration fit) and hence slightly differs from the classical variance-based sensitivity analysis. 

We observe that the calibrated $\lambda$'s can take quite small values, but as was shown in \cite{Campolongo06}, even in that case, the jumps might effect option prices significantly, especially for out-of-the-money contracts. 

\begin{table}
\centering
{\small
\caption{Importance of $\lambda$ for calibrations Apple Inc. stock on all four datasets. Hypothesis $0$ denotes we were unable to reject the null hypothesis and vice versa for $1$.}
\label{Tab:SenLambda}
\begin{tabular}{lllll} 
Data sets & 1/4/2015 & 15/4/2015 & 1/5/2015 & 15/5/2015 \\
 \midrule 
Hypothesis & $1$ & $1$ & $1$ & $1$\\
p-value & $1.30\%$ & $0.43\%$ & $8.45\text{e-}12\%$ & $3.56\%$ \\
 \bottomrule
\end{tabular}
}
\end{table}

\begin{figure}
\begin{subfigure}[b]{0.5\textwidth}
\includegraphics[width=\textwidth]{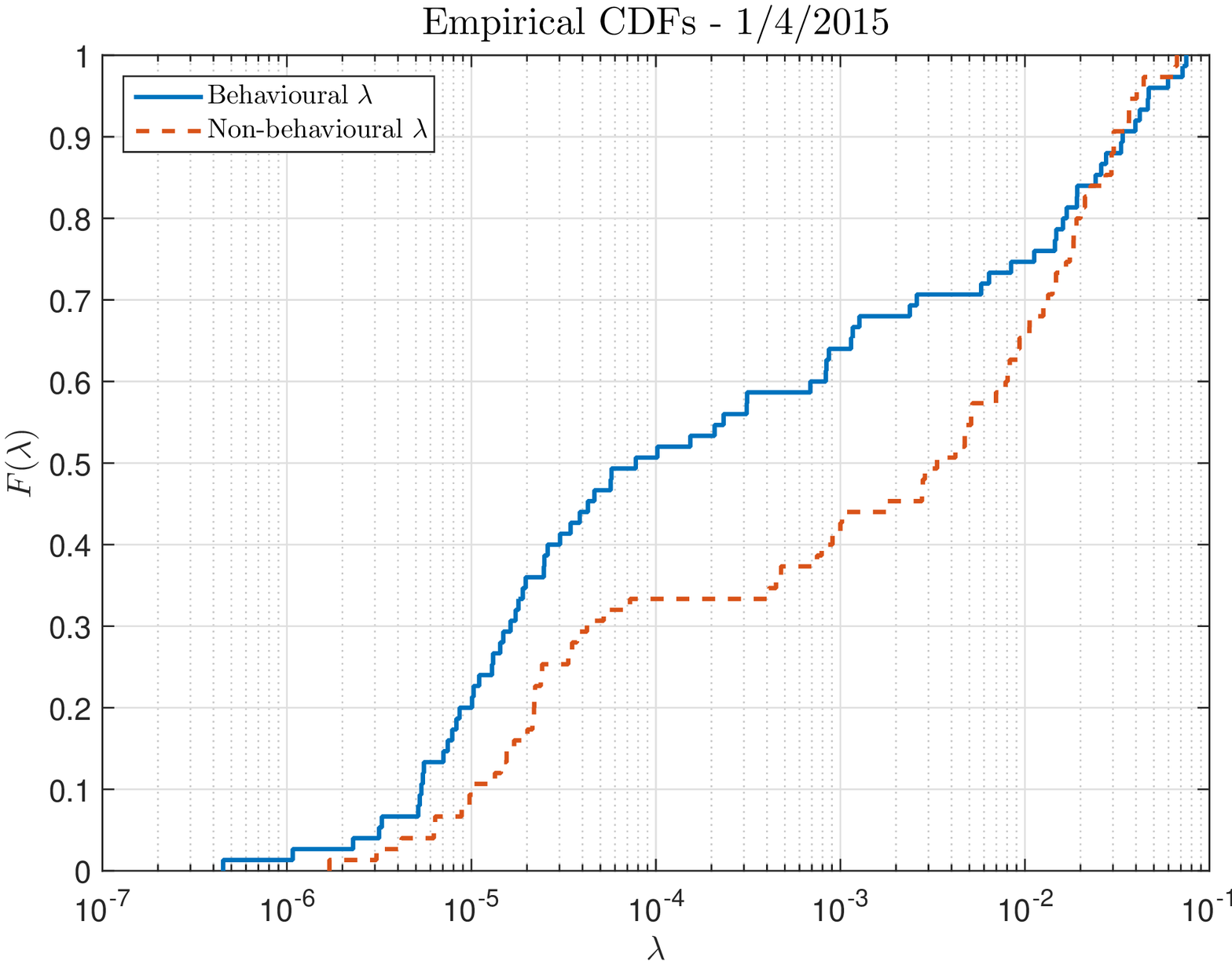}
\end{subfigure}
\begin{subfigure}[b]{0.5\textwidth}
\includegraphics[width=\textwidth]{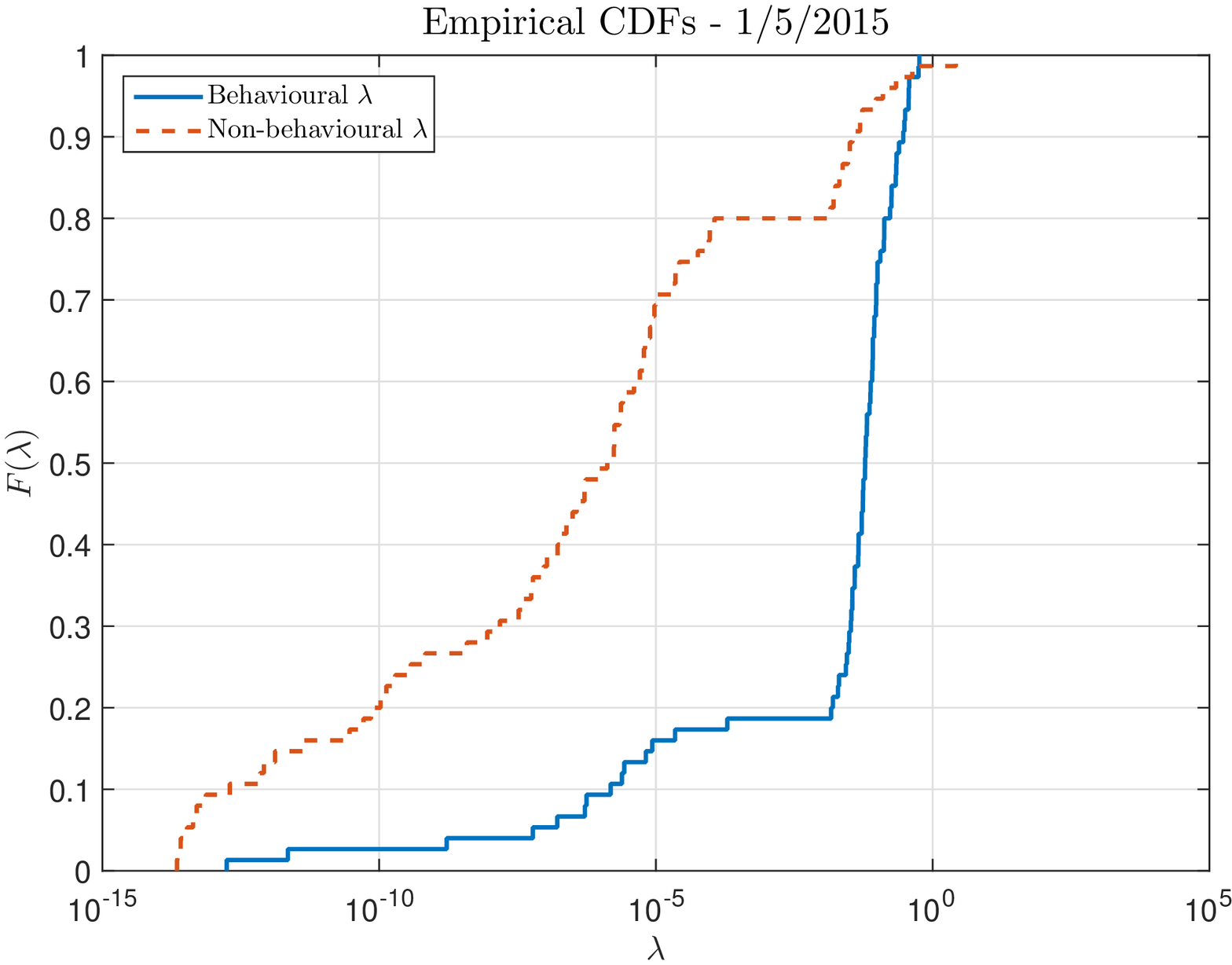}
\end{subfigure}
\caption{Empirical cumulative distribution functions of both sets ($\lambda$).}
\end{figure}

\subsection*{Sensitivity of the calibration with respect to the Hurst parameter}

Following the procedure of Monte-Carlo filtering for jump intensity $\lambda$ we are interested in the importance of the Hurst parameter. Since for $H=0.5$ one gets
a standard stochastic volatility model with jumps, if we are able to conclude that calibration of $H$ is crucial to obtain a good market fit, then we get a
justification of the approximative fractional model which is in-line with the long-memory phenomenon of realized volatility time series.

We were able to reject the null hypothesis for data sets from $1^{st}$ April and May and also from $15^{th}$ May. P-values were quite small (see Table
\ref{Tab:SenH}) for these data sets, unlike for the data from $15^{th}$ April. In this case were not able to reject the null hypothesis at any reasonable level of significance and hence we cannot make any conclusion regarding this data set.

\begin{table}[ht]
\centering
{\small
\caption{Importance of $H$ for calibrations Apple Inc. stock on all four datasets. Hypothesis $0$ denotes we were unable to reject the null hypothesis and vice versa for $1$.}
\label{Tab:SenH}
\begin{tabular}{lllll} 
Data sets & 1/4/2015 & 15/4/2015 & 1/5/2015 & 15/5/2015 \\
 \midrule 
Hypothesis & $1$ & $0$ & $1$ & $1$\\
p-value & $2.78\text{e-}10\%$ & $30.00\%$ & $5.08\text{e-}03\%$ & $2.17\%$ \\
 \bottomrule
\end{tabular}
}
\end{table}

\begin{figure}
\begin{subfigure}[b]{0.5\textwidth}
\includegraphics[width=\textwidth]{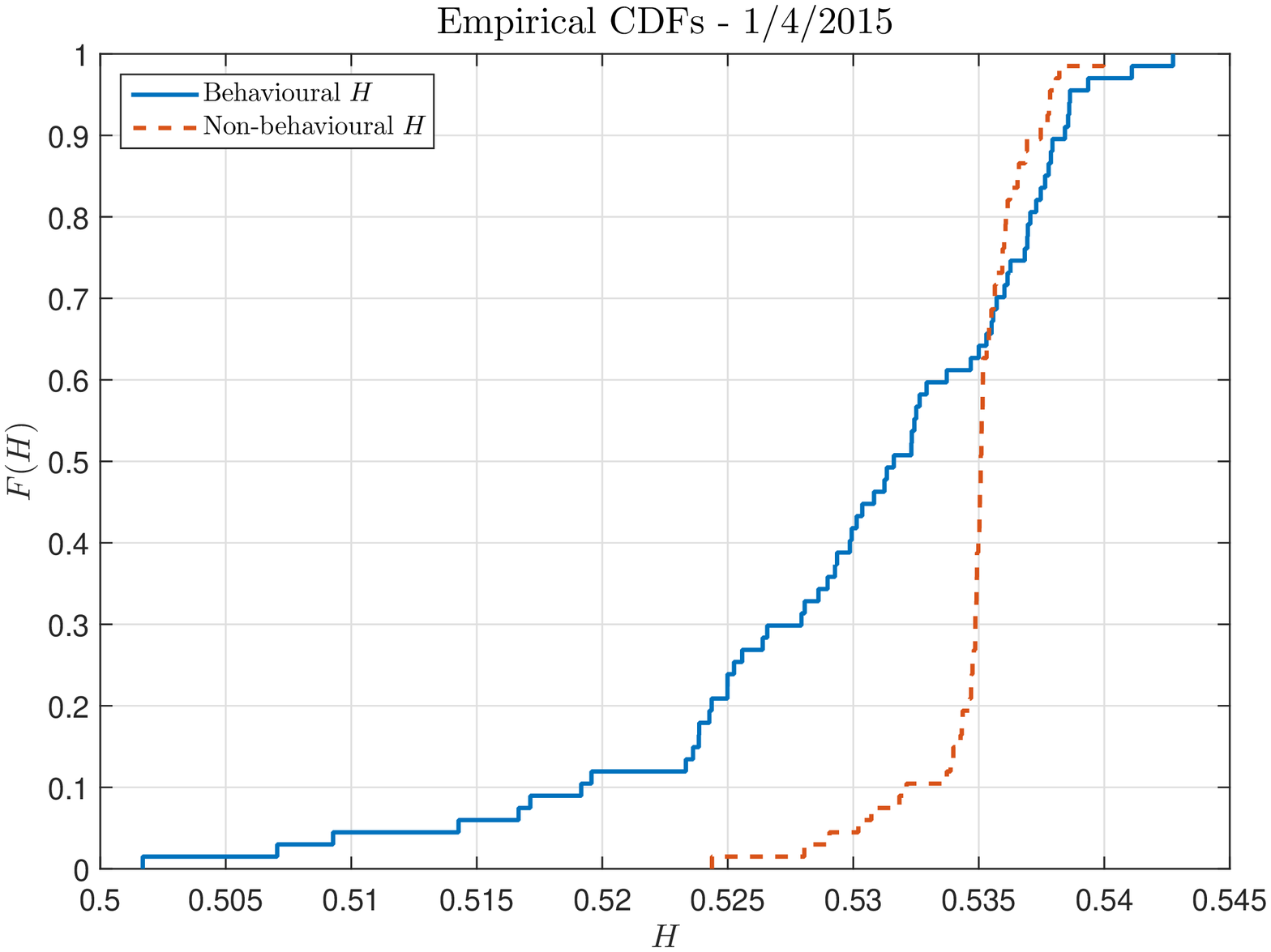}
\end{subfigure}
\begin{subfigure}[b]{0.5\textwidth}
\includegraphics[width=\textwidth]{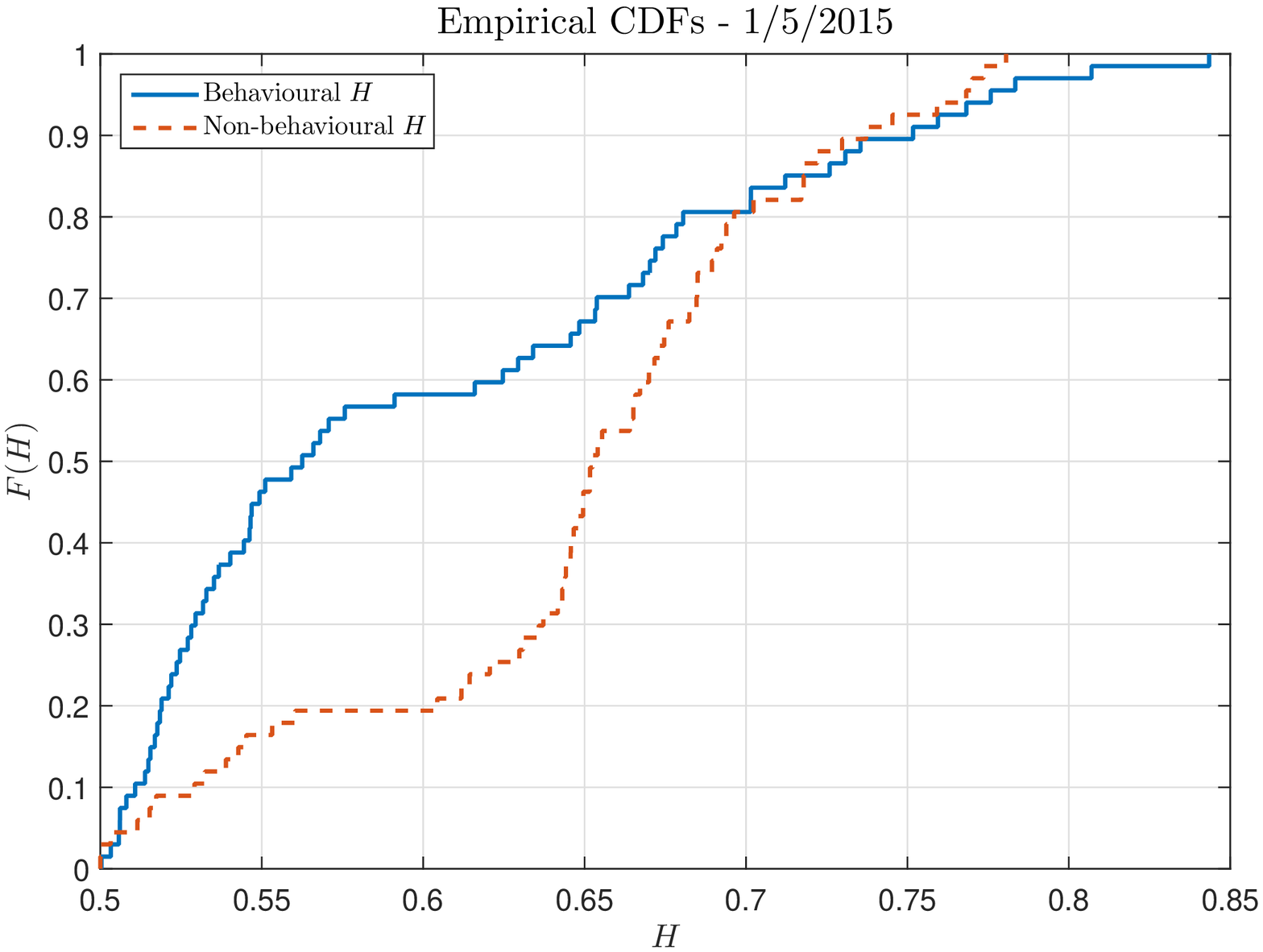}
\end{subfigure}
\caption{Empirical cumulative distribution functions of both sets ($H$).}
\end{figure}

\section{Conclusion}\label{sec:conclusion}

In this paper we performed the robustness and sensitivity analysis of several continuous-time stochastic volatility models (Heston, Bates and FSV model) with respect to market calibration. Using the bootstrap method we calibrated the model parameters 200 times and we compared all three models with respect to the variation in model parameters and in bootstrapped option prices. 

The bootstrap relative errors of all three models (Figure~\ref{Fig:C_Theta_pic}, data from 1/4/2015) are qualitatively similar -- the best errors are achieved by FSV model and the worst results by Heston model for all data sets. One can observe higher errors for OTM options ($K > S_0$). As for the bootstrap variances, the structure remains similar for all three models, but qualitatively the results differ strongly. Option prices (and hence market errors) obtained by Bates model have the largest variance with respect to changing data structure. Therefore, the Bates model appears to be the least robust. For 1/4/2015 data set, we retrieved the best bootstrap errors and lowest variances thereof by the FSV model. The Heston model can achieve lower variances (e.g. 15/5/2015), but bootstrap errors were greater compared to the FSV approach. 
From scatterplot matrices depicted in Figures \ref{Fig:ScatterBates} and \ref{Fig:ScatterFSV} we can observe that the histograms of Bates and FSV model differ especially for parameters $\kappa$ and $\sigma$. It is worth to mention that considering different parameter bounds (cf. Figure~\ref{Fig:Structure}b) may lead to different calibration results, with values of some of the calibrated parameters close to the boundary. Since $\kappa$ is the parameter with the most correlations, we performed all the tests with relatively high upper bound ($\kappa\leq 100$).  In the scatterplot matrices one can further see non-statistical connections of jump parameters, especially for the Bates model. To avoid this, one could fix one jump parameter for the calibration process (e.g. $\lambda$).

In Figure \ref{Fig:QQHeston} we can observe that the calibrated parameters for the Heston model are almost normally distributed unlike for the other models.  One should be careful especially with normality assumptions of $\Theta$. Additionally the calibrated parameters cannot be modelled as independent random variables (see Figures \ref{Fig:ScatterBates} and \ref{Fig:ScatterFSV}), therefore standard sensitivity analysis tests are not suitable in this context.  For this reason we used the Monte-Carlo filtering technique to show the importance of the jumps intensity $\lambda$ in the Bates model and the importance of long memory parameter $H$ in the FSV model. As for the jumps, in all four considered data sets we were able to conclude that considering jumps (non-zero $\lambda$) in a model plays a significant role in calibration to real market data. Even small values of $\lambda$ can effect the call prices, especially for the out-of-the-money contracts.  We could say that calibration of the fractionality parameter $H$ is important only in three cases out of four.

Recently, \cite{MrazekPospisilSobotka16ejor} studied the calibration task for FSV model and compared it to the Heston case with respect to in- and out-of-sample errors on equity index data sets. Our study confirms that the approximative fractional model can outperform other studied SV models (see Table \ref{Tab:Overallerror}). Moreover, we show that this approach is more robust with respect to the uncertainty in the data structure, especially when compared to the other jump-diffusion model. However, it is surprising that an additional parameter (Hurst parameter $H$) lead to smaller bootstrap variance. We are also able to draw the conclusion that jumps can also lead to decreased robustness (Bates model), so the importance of jump terms discussed in \cite{Campolongo06} can effect model performance in a negative manner as well. 

\subsection{Further issues}
As mentioned in the introduction, we have considered only a long-memory regime ($H>0.5$) of the FSV model, due to technical restrictions of the pricing solution. \cite{Bayer15} have shown that a simple rough paths volatility model can perform surprisingly well even for short maturities, unlike the standard diffusion volatility models without jumps. This observation was also supported by \cite{Fukasawa11}, who has shown a jump-like behaviour of the rough volatility model. Incorporating a rough volatility regime ($H<0.5$) could also improve robustness of the model in terms of criteria introduced in this paper. Verification of this hypothesis is still due to a further research. In fact, the proposed methodology can be successfully applied to a rough volatility model as soon as one has an efficient pricing solution.

\section*{Funding}

\noindent This work was supported by the GACR Grant 14-11559S Analysis of Fractional Stochastic Volatility Models and their Grid Implementation. 

\section*{Acknowledgements}

\noindent Computational resources were provided by the CESNET LM2015042 and the CERIT Scientific Cloud LM2015085, provided under the programme "Projects of Large Research, Development, and Innovations Infrastructures".

\section*{Conflict of interest}

\noindent The authors declare that they have no conflict of interest.

\section*{Ethical approval}

\noindent This article does not contain any studies with human participants or animals performed by any of the authors.
\clearpage

{

}

\end{document}